\documentclass[aps,superscriptaddress]{revtex4}
\usepackage{amssymb}
\usepackage{amsmath}

\usepackage{graphicx}
\usepackage{dcolumn}

\def\be {\begin{equation}}
\def\ee  {\end{equation}}
\def\bea {\begin{eqnarray}}
\def\eea {\end{eqnarray}}

\newcommand{\hrad}{\hat{H}_{\mathrm{rad}}}

\begin{document}


\title{Quantum gravity and the Coulomb potential}

\author{Viqar Husain}\email{husain@math.unb.ca}
\affiliation{Department of Mathematics and Statistics,\\
University of New Brunswick, Fredericton, NB E3B 5A3, Canada}
\affiliation{Perimeter Institute for Theoretical Physics,\\
31 Caroline Street North, 
Waterloo, N2L 2Y5 ON, Canada}
\author{Jorma Louko}\email{jorma.louko@nottingham.ac.uk}
\affiliation{School of Mathematical Sciences,
University of Nottingham,
Nottingham NG7 2RD, United Kingdom}
\author{Oliver Winkler}\email{owinkler@perimeterinstitute.ca}
\affiliation{Department of Mathematics and Statistics,\\
University of New Brunswick, Fredericton, NB E3B 5A3, Canada}
\affiliation{Perimeter Institute for Theoretical Physics,\\
31 Caroline Street North, 
Waterloo, N2L 2Y5 ON, Canada}


\thispagestyle{empty}

\vspace{.5ex} 

\date{Revised September 2007. 
Published in Phys.\ Rev.\ D {\bf 76}, 084002 (2007).}

\begin{abstract}

We apply a singularity resolution technique utilized in loop quantum
gravity to the polymer representation of quantum mechanics on
$\mathbb{R}$ with the singular $-1/|x|$ potential.  On an equispaced
lattice, the resulting eigenvalue problem is identical to a finite
difference approximation of the Schr\"odinger equation. We find
numerically that the antisymmetric sector has an energy spectrum that
converges to the usual Coulomb spectrum as the lattice spacing is
reduced. For the symmetric sector, in contrast, the effect of the
lattice spacing is similar to that of a continuum self-adjointness
boundary condition at $x=0$, and its effect on the ground state is
significant even if the spacing is much below the Bohr radius.
Boundary conditions at the singularity thus have a significant effect
on the polymer quantization spectrum even after the singularity has
been regularized.

\end{abstract}


\maketitle


\section{Introduction}
\label{sec:intro} 

It is expected that a viable quantum theory of gravity will have to
say something about what happens to the curvature singularities in
classical solutions of general relativity. This would involve making
concrete intuitive ideas about the role to be played by the Planck
length as a fundamental discreteness scale. A possible guide
concerning how to do this is the quantum resolution of the Coulomb
potential in quantum mechanics, where the basic result is that the
expectation value $\langle \widehat{1/r} \rangle$ is finite in all
eigenstates of the Hamiltonian. This kinematic result together with
dynamical Coulomb scattering may be taken to constitute quantum
singularity avoidance associated with the 
classically singular $-1/r$ potential.

In the Hamiltonian approach to quantum gravity, the problem of
quantization from a mathematical point of view is to find a suitable
representation of an algebra of functions of position and momentum as
operators on a Hilbert space.  The Wheeler-DeWitt approach utilizes a
functional Schrodinger representation where the basic variables are
the Arnowitt-Deser-Misner (ADM) variables
\cite{dewitt-can,wheeler-can}. The loop quantum gravity (LQG) approach
uses a Poisson algebra of functions of a connection and triad, based
on loops and surfaces, to build a quantum
theory~\cite{thiemann-book}. The basic variables quantized are the
holonomy of the connection along a curve, and the integral of the
(densitized) triad over a surface. The latter approach naturally leads
to a non-separable kinematical 
Hilbert space associated with graphs embedded in a
spatial manifold~\cite{thiemann-book}. Because of the association of
kinematic states with graphs, there is an intrinsic spatial discreteness
built into the quantum theory which is not a priori present in the
Schr\"odinger approach. Field excitations of the basic operators are
probed on graphs rather than at points.

An approach similar to that of loop quantum gravity can be employed
for the quantum mechanics of a particle moving in a potential in one
or more dimensions. Here graphs are replaced by lattices of spatial
points (not necessarily equispaced), and the basic observables
realized in the quantum theory are the configuration and translation
operators. This is of course natural since 
generators of infinitesimal translations cannot be 
represented on a spatial lattice. For a
given equispaced lattice, the position eigenstates are normalizable,
and the Hilbert space is obviously separable. However, the Hilbert
space that incorporates all possible lattices (equispaced or not) is
non-separable. The quantum theory that utilizes this space has been
referred to as polymer quantization~\cite{polymer}.

From a mathematical viewpoint, conventional Schr\"odinger quantization
and polymer quantization are inequivalent. In the former, wave
functions are square-integrable functions on~$\mathbb{R}^3$, with
position and momentum operators acting as multiplication and
differential operators, respectively.  A~remarkable fact is that this
is quantization is unique up to unitary equivalence provided that the
configuration space of the system is topologically $\mathbb{R}^3$
and that the representation of the Weyl-Heisenberg algebra of
exponentiated position and momentum operators is weakly
continuous~\cite{polymer}. It is known, however, that if either of
these assumptions is abandoned, there are infinitely many inequivalent
representations. A~simple example is a system where the configuration
space is the torus rather than~$\mathbb{R}^3$. Polymer quantization
provides another example. Its non-separable Hilbert space may be
viewed as the inductive limit of the separable Hilbert spaces
associated with quantum mechanics on all possible lattices, including
the one-parameter family of equispaced ones.

From a practical viewpoint, since only a finite number of
calculations are possible, the full polymer Hilbert space can never be
utilized. Rather, only a separable subspace is computationally
useful. In this case, however, polymer quantization appears to reduce
to the finitely-differenced Schr\"odinger equation on a lattice. We
elaborate on this below by observing that various finite difference
schemes for differential equations may be rewritten using
configuration and translation operators on a lattice.

In the representation used in LQG, there is a way to write inverse
triad operators using certain classical Poisson bracket identities due
to Thiemann~\cite{thomas}. 
These identities are in fact much more
general than the context in which they first arose; similar ones may
be written for any theory on a lattice. They may be used to write
inverse scale factor operators that are bounded on kinematical states
in quantum cosmology in both the connection-triad and ADM variables
\cite{mb1,hw-cosm}, as well as curvature operators for a field theoretic 
formalism for gravitational collapse \cite{hw-bh}. The boundedness
property of such operators has been used in discussions of singularity
avoidance in quantum gravity at the kinematical
level~\cite{th-br1,th-br2}.

In this paper we apply these ideas of singularity avoidance in quantum
gravity to polymer representation quantum mechanics on $\mathbb{R}$
with the singular $-1/|x|$ potential. For $x>0$, and with an
appropriate boundary condition at $x=0$, this may be thought of as the
spherically symmetric sector of the Coulomb problem
on~$\mathbb{R}^3$. We address two main questions. First, we show that
in any lattice context, the LQG singularity resolution technique is
equivalent to replacing a singular derivative by a non-singular finite
difference scheme. Second, we show that even after the singularity in
the potential has been resolved, the spectra in the symmetric and
antisymmetric sectors are significantly different. The latter is close
to the usual Coulomb spectrum when the lattice spacing is much below
the Bohr radius, but in the former the lattice spacing plays a role
similar to a continuum self-adjointness boundary condition at $x=0$,
and the effect on the ground state is significant even when the
spacing is much below the Bohr radius. We conclude that boundary
conditions at the singularity have a significant effect on the polymer
quantization spectrum even after the singularity itself has been
regularized.

The rest of the paper is as follows: In section \ref{sec:polymer} we
recall the basic structure of polymer quantization on~$\mathbb{R}$. In
section \ref{sec:radialcoulomb} we specialize to the
potential~$-1/|x|$, introducing the lattice regularization of the
potential and showing that the boundary conditions of the radial
Coulomb problem can be implemented by the restriction to the
antisymmetric sector. Our numerical results for the spectrum are given
in section~\ref{sec:spectrum}. The symmetric sector is analyzed in
section~\ref{sec:symmetric}. Section \ref{sec:discussion} summarizes
the results and discusses their implications and limitations for the
problem of singularity resolution in quantum gravity.


\section{Polymer quantization on $\mathbb{R}$}
\label{sec:polymer}

In this section we briefly describe how a mechanical system is
quantized in the polymer representation~\cite{polymer}. To keep the
notation simple, and because the radial Coulomb problem we discuss
later is a one-dimensional system, we will consider a particle on the
real line. The generalization to $n$ particles in $\mathbb{R}^3$ is
straightforward.

Recall that the Hilbert space for the Schr\"odinger quantization of a
particle on the real line is $L_2(\mathbb{R})$, the space of
square-integrable functions on $\mathbb{R}$ in the Lebesgue
measure. The operators corresponding to configuration and momentum
variables act respectively as multiplication and differentiation
operators. The Hilbert space is separable; an example of a countable
basis are the harmonic oscillator eigenfunctions.

To introduce polymer quantization on~$\mathbb{R}$, we start with the
basis states 
\be 
\psi_{x_0}(x) = \left \{ \begin{array}{r l} 1, & x=x_0 \\
0, & x\ne x_0 \ . \end{array} 
\right . 
\label{eq:basisstates}
\ee 
The polymer Hilbert space is the Cauchy completion of the linear span
of these basis states in the inner product 
\be 
\langle \psi_x , \psi_{x'} \rangle = \delta_{x,x'}, 
\label{eq:ip}
\ee 
where the quantity on the right-hand side 
is the Kronecker (rather than Dirac) delta. 
This space is clearly nonseparable, and hence 
inequivalent to $L_2(\mathbb{R})$ 
\cite{velhinho,corduneau,hewitt-ross}. Intuitively,
building from the polymer basis states a single nonzero 
$L_2(\mathbb{R})$ 
state
would require an
uncountable superposition, and thus lead to an unnormalizable state in
the polymer Hilbert space. 
Conversely, any state in the polymer Hilbert space
has support on at most countably many points, and
will thus represent the zero state in $L_2(\mathbb{R})$. 

Next we define the action of the basic quantum operators. 
The position operator 
$\hat{x}$ acts by multiplication, 
\be 
\bigl(\hat{x} \psi \bigr) (x) =  x \psi(x), 
\label{eq:xhat-action}
\ee 
and its domain contains the linear span of the basis
states~(\ref{eq:basisstates}). 
The translation operators $\hat{U}_{\lambda}$, 
$\lambda\in\mathbb{R}$, act by 
\be 
\bigl( \hat{U}_{\lambda} \psi \bigr) (x)  
= \psi(x + \lambda) , 
\label{eq:Uhat-action}
\ee 
and are clearly unitary. 
Formulas (\ref{eq:xhat-action}) and (\ref{eq:Uhat-action}) are
identical to those in $L_2(\mathbb{R})$. In $L_2(\mathbb{R})$, 
the action of $\hat{U}_{\lambda}$ is
weakly continuous in~$\lambda$, and there exists a densely-defined
self-adjoint momentum operator $\hat{p}$ such that 
$\hat{p} = -i
\bigl[\partial_\lambda
\hat{U}_{\lambda}\bigr]_{\lambda=0} 
= -i \partial_x$ and $\hat{U}_{\lambda}
= e^{i\lambda \hat{p}}$. By contrast, in the polymer Hilbert space the
action of $\hat{U}_{\lambda}$ is not weakly continuous in~$\lambda$,
and a basic momentum operator does not exist.

The states in the polymer Hilbert space can be described as points in a
certain compact space, the (Harald) 
Bohr compactification of the real line, and
the operators introduced above can be described in terms of a
representation of the Weyl algebra associated with the classical
position and momentum variables
\cite{velhinho,corduneau,hewitt-ross}. There exists also a
mirror-image quantization in which a a momentum operator and a family
of translation operators in the momenta exist, but there is no basic
position operator~\cite{halvorson}. These mathematical structures will
however not be used in the rest of the paper.

As there is no basic momentum operator, any phase space
function containing the classical momentum~$p$, most importantly the
Hamiltonian, has to be quantized in an indirect way. 
Following~\cite{polymer}, we fix a length scale $\lambda>0$ and define 
\begin{subequations}
\label{eq:p-and-psquared}
\begin{align}
\hat{p} &= \frac{1}{2i \lambda} (\hat{U}_\lambda -
\hat{U}^{-1}_{\lambda} ) , 
\\
\widehat{p^2} &= 
\frac{1}{\lambda^2} ( 2- \hat{U}_\lambda -
\hat{U}^{\dagger}_{\lambda} ) . 
\end{align} 
\end{subequations}
The Hamiltonian operator that corresponds to the classical Hamiltonian 
$H = \frac12 p^2 + V(x)$ is then 
\be 
\hat{H} =
\frac{1}{2\lambda^2} ( 2- \hat{U}_\lambda -
\hat{U}^{\dagger}_{\lambda} ) + \hat{V} ,
\label{H}
\ee 
where $V$ is assumed so regular that 
$\hat{V}$ can be defined by pointwise multiplication, 
$\bigl(\hat{V} \psi \bigr) (x) =  V(x) \psi(x)$. 
In $L_2(\mathbb{R})$, the $\lambda\to0$ limit in 
(\ref{eq:p-and-psquared}) would give the usual 
momentum and momentum-squared operators 
$-i\partial_x$ and~$-\partial^2_x$, and the kinetic term in 
(\ref{H}) would reduce to $-\frac12\partial^2_x$. 
In the polymer Hilbert space the $\lambda\to0$ limit does not exist, 
and $\lambda$ is regarded as a fundamental length scale. 
For $\lambda \ll 1$, 
one expects the polymer dynamics to be well approximated by the
Schr\"odinger dynamics, and certain results to this effect are known
\cite{polymer,fredenhagen-reszewski,corichi-vuka-zapata}.


Although the polymer Hilbert space is nonseparable, the dynamics
generated by $\hat{H}$ (\ref{H}) breaks into separable superselection
sectors. To discuss this, it is convenient to introduce a Dirac
bra-ket notation in which the basis state $\psi_{\mu}$
(\ref{eq:basisstates}) is denoted by~$|\mu \rangle$. From
(\ref{eq:ip}), (\ref{eq:xhat-action}) and (\ref{eq:Uhat-action}) we
then have
\begin{align}
\langle \mu | \mu' \rangle
& = \delta_{\mu,\mu'} , 
\\
\hat{x} \, |\mu \rangle 
&= \mu |\mu\rangle , 
\\
\hat{U}_\lambda |\mu \rangle 
&= |\mu - \lambda \rangle .
\end{align}
The action of $\hat{H}$ on $|\mu\rangle$ gives a state with support at
$\mu$, $\mu-\lambda$ and $\mu+\lambda$. The time evolution of
$|\mu\rangle$ thus has support only on the regular $\lambda$-spaced
lattice $\bigl\{\mu + n\lambda \mid n\in\mathbb{Z} \bigr\}$. This
means that the time evolution breaks into superselection sectors,
where each sector has support on a regular $\lambda$-spaced lattice
and is hence separable, and the sectors can be labelled by the lattice
point $\mu \in [0, \lambda)$. The time evolution of any given initial
state will consequently have support only on a countable union of
$\lambda$-spaced lattices. The upshot is that the time evolution of
any separable subspace is restricted to a separable subspace. Thus,
even though the polymer Hilbert space is nonseparable, the fundamental
length scale $\lambda$ and the choice of an initial state or an
initial separable subspace will result in quantum dynamics that
takes place in a separable Hilbert space.

We note that on a fixed $\lambda$-spaced lattice, the Hamiltonian 
(\ref{H}) agrees with a conventional 
discretisation of Schr\"odinger's equation by the replacement 
\be 
\psi''(x_n) \rightarrow \frac{1}{\lambda^2}\, 
\left( \psi_{n+1}-2\psi_n +\psi_{n-1}\right) . 
\ee
This suggests investigating versions of 
(\ref{H}) in which the kinetic term is replaced by an operator that, 
in the finite difference approximation context, 
is higher-order accurate in~$\lambda$. 
The discussion of superselection sectors would extend 
to such versions with only minor changes. 
As the main interest in the present paper concerns 
singular potentials rather than higher order accurate discretizations of 
the second derivative, we shall work with~(\ref{H}). 

In summary, the restriction of the polymer dynamics into any of its
superselection sectors is mathematically equivalent to a conventional
discrete approximation to the continuum Schr\"odinger equation on the
corresponding equispaced lattice. The conceptual difference is, however,
that in the polymer theory the lattice spacing is regarded as a new
fundamental scale.


\section{The radial Coulomb problem on a lattice}
\label{sec:radialcoulomb}

Reduction of Schr\"odinger's equation with the Coulomb potential to
the spherically symmetric ($l=0$) sector yields the radial Hamiltonian
operator
\be 
\hrad = 
-\frac{d^2}{dr^2} -
\frac{1}{r} , 
\label{Hyd} 
\ee 
acting in the Hilbert space of square integrable functions on 
$(0,\infty)$ in the measure $dr$
\cite{merzbacher}. (For numerical convenience, the radial coordinate
$r$ has been chosen as twice the Rydberg radial coordinate.) $\hrad$
has a one-parameter family of self-adjoint extensions, each
characterized by a boundary condition at
$r\to0$~\cite{fewster-hydrogen}. The conventional choice for the 
self-adjoint extension is to assume the 
three-dimensional eigenfunctions in $L_2(\mathbb{R}^3, d^3\mathbf{x})$
to be bounded at the origin: in terms of the rescaled wave functions
on which $\hrad$ acts, this means that the wave function vanishes at
$r\to0$~\cite{merzbacher}. The spectrum then consists of the positive
continuum and the negative discrete eigenvalues
\be \epsilon_n = -\frac {1}{4n^2}, \ \ n = 1,2,\ldots \ . 
\label{eq:continuum-En} 
\ee
We shall return to the other possible choices of the self-adjointness
boundary condition in section~\ref{sec:symmetric}.

To introduce a polymer counterpart of $\hrad$ along the lines
of~(\ref{H}), we need to address the positivity of~$r$, the boundary
condition at $r=0$, and the singularity of the potential at
$r\to0$. Suppressing the singularity issue for the moment, the first
two issues can be solved by extending $r$ to negative values: denoting
this extended coordinate by $x\in\mathbb{R}$, and assuming that the
discretized potential is symmetric under the reflection $x\mapsto-x$,
we require the states to be antisymmetric under $x\mapsto-x$. In the
two superselection sectors in which the lattice points are at
respectively $x_n = n\lambda$ and $x_n = (n+\frac12)\lambda$,
$n\in\mathbb{Z}$, this antisymmetry condition just chooses the
antisymmetric states. The remaining superselection sectors are
pairwise coupled by the antisymmetry condition. However, since the
potential is by assumption symmetric under $x\mapsto-x$, the energy
eigenvalues can be found within each sector without using an
antisymmetry condition, and the antisymmetric eigenstates are then
obtained by just taking appropriate linear combinations.

Let us return now to the singularity of the potential at
$x=0$. Although this issue only arises in the single superselection
sector that has a lattice point at $x=0$, we wish to give a
prescription that handles all the superselection sectors in a unified
manner. Let $x_n$ denote the lattice points. Since
\be
\frac{\mathrm{sgn}(x)}{\sqrt{|x|}} = 2 \frac{d(\sqrt{|x|})}{dx}  ,
\label{eq:diff-identity}
\ee 
we can represent 
$\bigl(\mathrm{sgn}(x)\bigr)/\bigl(\sqrt{|x|}\bigr)$ 
by a finite-difference version of the derivative, 
\be
\frac{\mathrm{sgn}(x_n)}{\sqrt{|x_n|}}
\to 
\frac{1}{\lambda} 
\left(\sqrt{|x_{n+1}|} - \sqrt{|x_{n-1}|} \right) . 
\ee
Taking the square, this leads to the lattice potential 
\be 
- \frac{1}{|x_n|} 
\rightarrow 
- \frac{1}{\lambda^2} 
\left(\sqrt{|x_{n+1}|}-\sqrt{|x_{n-1}|}\right)^2 , 
\label{findif1/x}
\ee
which is well defined even for $x_n=0$. The resulting Hamiltonian
operator can be written in terms of the fundamental translation and
multiplication operators as
\begin{align}
\hat{H} 
&= 
\frac{1}{\lambda^2} \left( 2- \hat{U}_\lambda -
\hat{U}_\lambda^\dagger\right)
- \frac{1}{\lambda^2} 
\left( 
\hat{U}_\lambda \sqrt{|x|}\, \hat{U}_\lambda^\dagger
- 
\hat{U}_\lambda^\dagger \sqrt{|x|}\, \hat{U}_\lambda
\right)^2 
\nonumber
\\
&= 
\frac{1}{\lambda^2} \left( 2- \hat{U}_\lambda -
\hat{U}_\lambda^\dagger\right)
- \frac{1}{\lambda^2} 
\left( \hat{U}_\lambda^\dagger\left[\hat{U}_\lambda, \sqrt{|x|}\
\right] - \hat{U}_\lambda\left[\hat{U}^\dagger_\lambda, \sqrt{|x|}\
\right]\right)^2 , 
\label{Hpoly}
\end{align}
and its action on the basis state $|\mu\rangle$ is 
\be \hat{H}|\mu\rangle 
= \frac{1}{\lambda^2}
\bigl( 2|\mu\rangle -
|\mu-\lambda\rangle - |\mu+\lambda\rangle 
\bigr) -
\frac{1}{\lambda^2}
\left( \sqrt{|\mu+\lambda|} -
\sqrt{|\mu-\lambda|} \right)^2 
|\mu\rangle . 
\label{hpolye}
\ee

\begin{figure}
\includegraphics[height=4in,width=5in,angle=0]{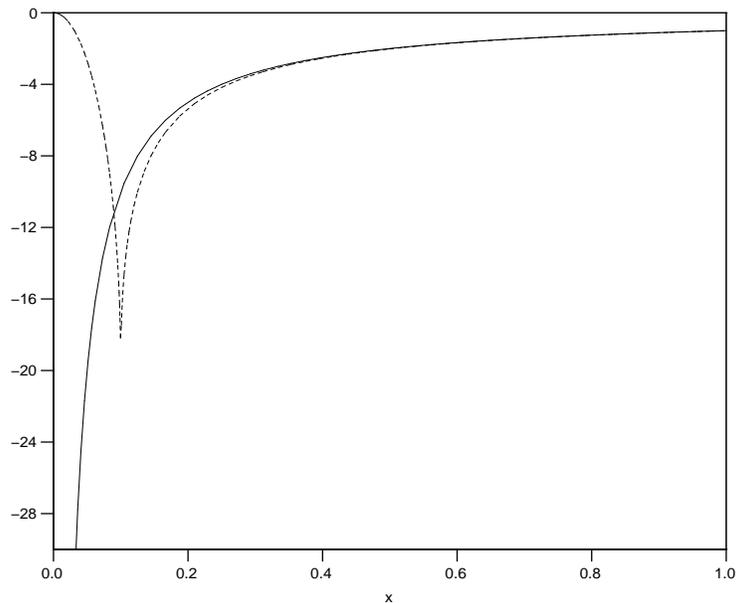}
 \caption{\baselineskip = 1.0em  The solid line is the Coulomb potential 
 and the dashed line is the lattice regularized potential 
(\ref{hpolye}) for $\lambda=0.1$  } \label{potentials}
\end{figure}

The potential term in the last form of (\ref{Hpoly}) 
could have been arrived at by considering a 
phase space version of the identity~(\ref{eq:diff-identity}), namely
\be 
\frac{\mathrm{sgn}(x)}{\sqrt{|x|}} 
= \frac{2}{i\lambda} e^{-i\lambda p} 
\left\{ \sqrt{|x|}, e^{i\lambda p} \right\} , 
\ee
where $x$ and $p$ are canonically conjugate variables. This would be
the route that led to Thiemann's regularisation of inverse triad
operators in LQG~\cite{thomas}. In the present context, we have
arrived at (\ref{Hpoly}) simply by representing a derivative by a
finite difference on a lattice. A~comparison of the Coulomb
potential and its lattice regulated version is shown in
Figure \ref{potentials} for $\lambda =0.1$.  There is a striking
repulsive modification near $x=0$, a result which in its quantum
gravity incarnation leads to a bounded curvature at the Planck
scale~\cite{hw-bh}.

%


\section{Spectrum}
\label{sec:spectrum} 

Let us focus now on the superselection sector with the lattice points
$x_n = n\lambda$, $\lambda \in \mathbb{Z}$. As this is the sector in
which resolving the singularity at $x=0$ is necessary, we expect it
will provide the most interesting test of the resolution
proposal~(\ref{findif1/x}).

We look for energy eigenstates in the form 
$\sum_n c_n | n\lambda \rangle$, 
where the coefficients $c_n$ are subject to the
normalizability condition $\sum_n {|c_n|}^2 < \infty$ 
and the antisymmetry condition $c_n = - c_{-n}$. The energy eigenvalue
equation with eigenvalue $\epsilon$ reads 
\be 
\hat{H}\sum_n c_n | n\lambda \rangle 
= \epsilon \sum_n c_n | n\lambda \rangle. 
\ee
It reduces to the recursion relation 
\be
c_n\left(2-\lambda f_n-\lambda^2 \epsilon\right) = c_{n+1} + c_{n-1},
\label{recr}
\ee
where
\be f_n = \left(\sqrt{|n-1|} - \sqrt{|n+1|}\right)^2.
\ee

Suppose from now on that $\epsilon<0$. 
The asymptotic form of the recursion relation (\ref{recr}) as
$n\to\infty$ is 
\be 
c_n(2- \lambda^2 \epsilon)=c_{n+1}+c_{n-1}. 
\label{recr-as}
\ee
It has the linearly independent solutions  
\be
c_n = \left[ 
1 - {\textstyle\frac12}\lambda^2 \epsilon  
+ 
\sqrt{ {\left( 1 - {\textstyle\frac12}\lambda^2 \epsilon \right)}^2 - 1}
\right]^{\pm n} . 
\label{eq:cn-as}
\ee
It follows \cite{elaydi} that the exact recursion relation
(\ref{recr}) has only one linearly independent solution that does not
grow exponentially as $n\to\infty$, and this solution has the
asymptotic form (\ref{eq:cn-as}) with the lower sign, and is hence
exponentially decreasing in~$n$. We use this observation to set up a
shooting method for a numerical computation of the eigenenergies. We
start at an initial $n_0 \gg 1/(-\epsilon\lambda)$, in which regime the
asymptotic recursion relation (\ref{recr-as}) holds, 
compute $c_{n_0 -1}$
using~(\ref{eq:cn-as}), and then compute $c_0$ using the exact
recurrence relation~(\ref{recr}). Because of the antisymmetry
condition $c_n = - c_{-n}$, the eigenenergies are those for
which $c_0=0$. The accuracy of the method is monitored by increasing
the value of $n_0$ until the results no longer change to the desired
accuracy.

When $\lambda = 0.1$, we find that the lowest 14 eigenvalues are such
that the quantity $k=1/\sqrt{-4\epsilon}$ is a few per cent below the  
lowest 14 positive integers, being thus a good approximation to the
continuum spectrum~(\ref{eq:continuum-En}). For the higher eigenvalues
the numerics becomes slow, and we do not have an estimate of when $k$
starts to differ significantly from integers.

\begin{figure}
\includegraphics[height=4in,width=5in,angle=0]{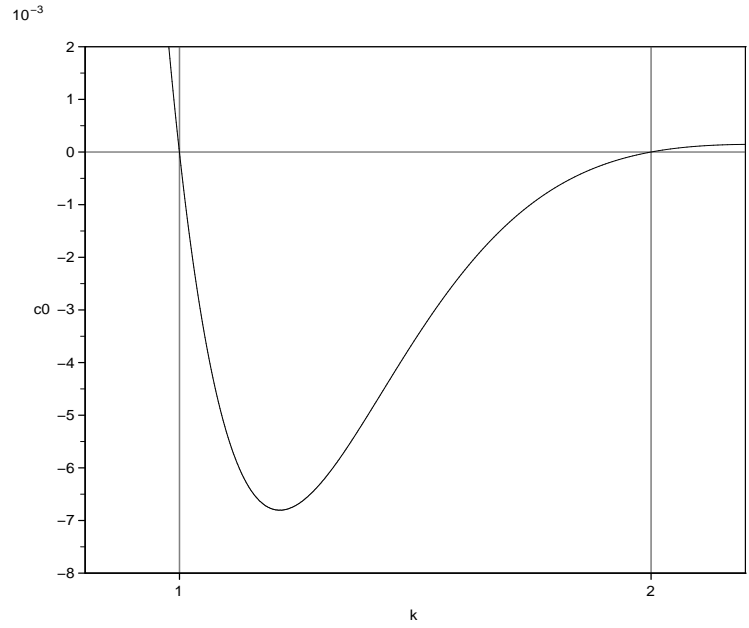}
\caption{\baselineskip = 1.0em  
The coefficient $c_0$ as a function of 
$k=1/\sqrt{-4\epsilon}$ for $0.98 \le k \le
2.2$, with $\lambda=0.01$. 
The zeroes are near $k=1$ and $k=2$.} 
\label{c012}
\end{figure}

\begin{figure}
\includegraphics[height=4in,width=5in,angle=0]{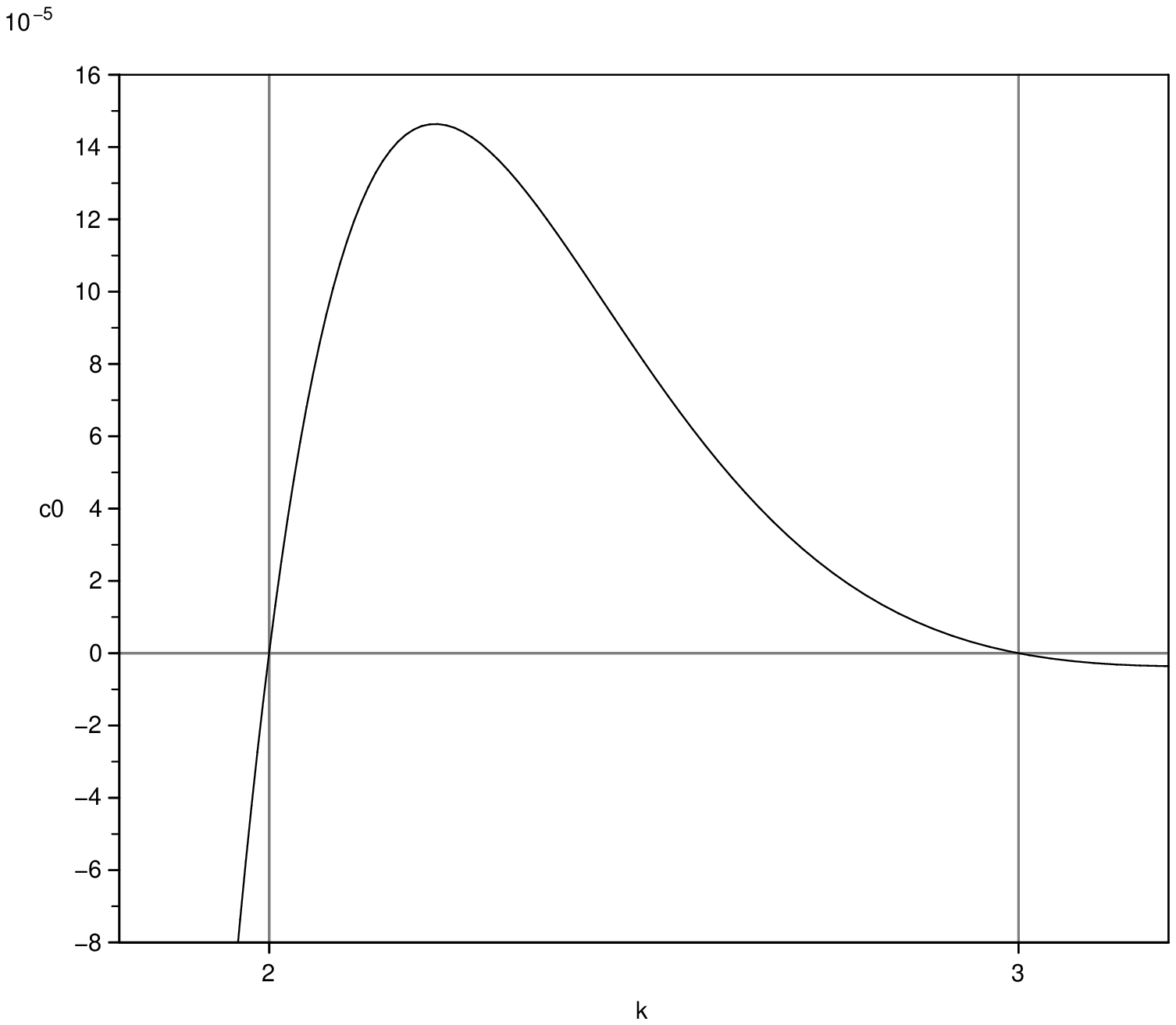}
\caption{\baselineskip = 1.0em  
The coefficient $c_0$ as a function of 
$k=1/\sqrt{-4\epsilon}$ for $1.98 \le k \le
3.2$, with $\lambda=0.01$. 
The zeroes are near $k=2$ and $k=3$.} 
\label{c023}
\end{figure}

\begin{figure}
\includegraphics[height=4in,width=5in,angle=0]{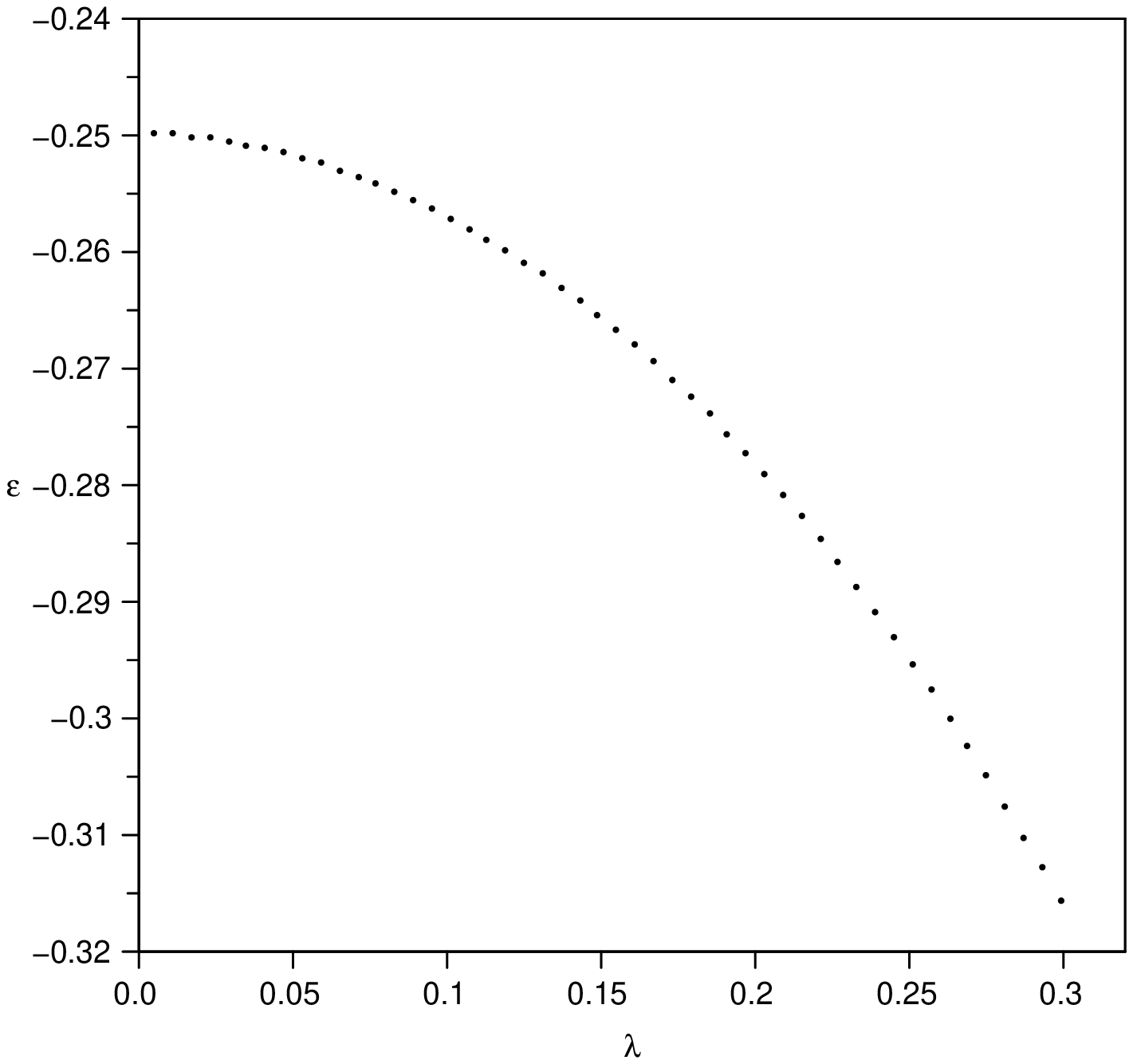}
\caption{\baselineskip = 1.0em 
The lowest eigenenergy as a function of $\lambda$ for $0.005 \le
\lambda \le 0.3$. The vertical error bar of each point is~$10^{-4}$. 
Convergence to $\epsilon=-0.25$ is apparent as $\lambda\to 0$.}
\label{e0lam}
\end{figure}

\begin{figure}
\includegraphics[height=4in,width=5in,angle=0]{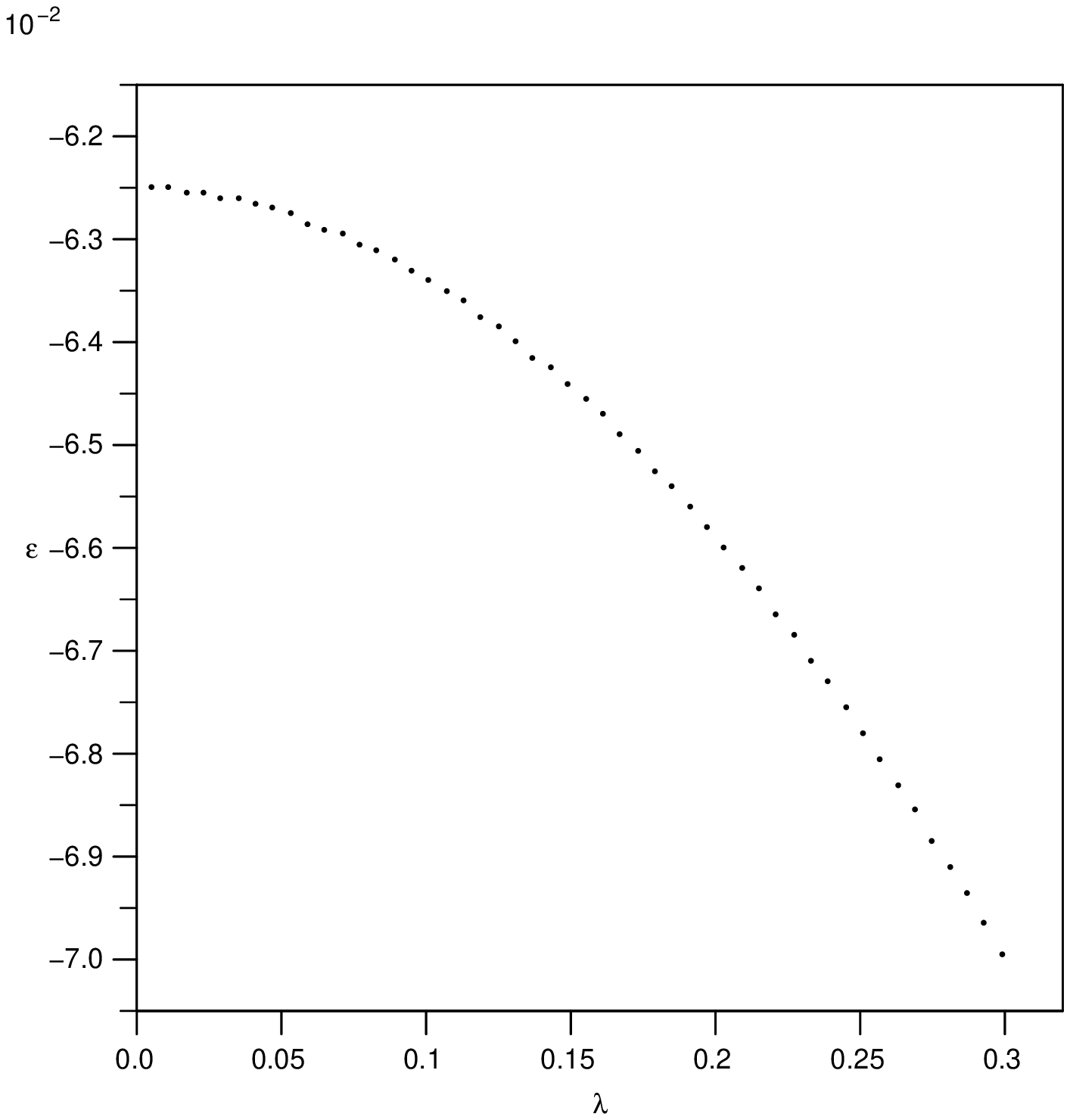}
\caption{\baselineskip = 1.0em  
The second-lowest eigenenergy as a function of $\lambda$ for $0.005 \le
\lambda \le 0.3$. 
The vertical error bar of each point is~$5\times 10^{-5}$. 
Convergence to 
$\epsilon=-0.0625$ is apparent as $\lambda \to 0$.}
\label{e1lam}
\end{figure}

For smaller $\lambda$ the numerics becomes slower but indicates
convergence towards the continuum eigenvalues from below as 
$\lambda\to0$. Figures
\ref{c012} and \ref{c023} show plots of 
$c_0$ as a function of $k$ for $\lambda=0.01$: within the resolution
of the plots, the three lowest roots are indistinguishable from $k=1$,
$k=2$ and $k=3$. Figures
\ref{e0lam} and \ref{e1lam} show 
the lowest two eigenenergies as functions of $\lambda$ for $0.005 \le
\lambda \le 0.3$.

%

\section{Symmetric boundary condition}
\label{sec:symmetric} 

The results in section 
\ref{sec:spectrum} 
show that the singularity resolution method (\ref{Hpoly}) gives good
agreement with the continuum results for the antisymmetric boundary
condition. While this is the boundary condition that arises from the
conventional treatment of the three-dimensional Coulomb
problem~\cite{fewster-hydrogen}, we now wish to consider the
resolution method for the $-1/|x|$ potential on the
\emph{full\/} real line in its own right. We must then find also the
energy eigenstates that are symmetric under $x\to-x$.

Staying in the superselection sector in which the lattice points are
$x_n = n\lambda$, $\lambda \in \mathbb{Z}$, the numerical algorithm 
of section 
\ref{sec:spectrum} 
can be adapted to the symmetric boundary condition by starting again
at $n_0 \gg 1/(-\epsilon\lambda)$ but computing now $c_1$
and~$c_{-1}$. The eigenenergies are those for which $c_1 = c_{-1}$.

The numerics is now considerably slower than for the antisymmetric
boundary condition. When $\lambda=0.1$, we find that there is a ground
state at $\epsilon \approx - 4.94$, well below the continuum hydrogen
ground state, and the first 12 excited states are such that the
quantity $k=1/\sqrt{-4\epsilon}$ is approximately 0.2 above the lowest
12 positive integers. When $\lambda$ decreases to $10^{-5}$, the
lowest eigenvalue decreases and shows no evidence of converging to a
limiting value as $\lambda\to0$, whereas the higher eigenvalues appear
slowly to converge from above to $-1/(4n^2)$,
$n=1,2,\ldots$. Eigenenergies of the ground state and the first
excited state for selected values of $\lambda$ are shown in
Table~\ref{symmtable}.

\begin{table}
\caption{\baselineskip = 1.0em  
The energy eigenvalue as a function of $\lambda$ for the ground state
$(\epsilon_0)$ 
and the first excited state $(\epsilon_1)$ 
with the symmetric boundary condition.}
\label{symmtable}
\begin{ruledtabular}
\begin{tabular}{ddd}
\lambda &-\epsilon_0 &-\epsilon_1
\\
\hline
10^{-1}& 4.94 & 0.153 
\\
10^{-2}& 14.8  & 0.181 
\\
10^{-3}& 32.3 & 0.196 
\\
10^{-4}& 58.5 & 0.207
\\
10^{-5}& 93.9 & 0.214
\end{tabular}
\end{ruledtabular}
\end{table}

The behaviour of the ground state in the limit $\lambda\to0$ has a 
counterpart in the continuum
theory. In the continuum theory, the solutions to the eigenvalue
differential equation $\hrad\psi = \epsilon\psi$ that do not vanish as
$r\to0_+$ have a logarithmic singularity there, and it is not possible
to single out a boundary condition at $r\to0_+$ by a `symmetric'
extension to negative $r$. Instead, the self-adjointness boundary
conditions at $r\to0_+$ can be parametrized by a length scale
$L\in\mathbb{R}
\cup \{\infty\}$, 
such that the eigenenergies are the solutions to the transcendental
equation 
\be
\frac{1}{L} = G \! \left( \frac{-1}{\sqrt{-4\epsilon}} \right), 
\ee
where
\be
G(z) := \Psi(1+z) - \ln|z| - 1/(2z)  
\ee
and $\Psi$ is the digamma function~\cite{fewster-hydrogen}. 
The usual boundary condition is obtained with $L=0$. For all the other
values of~$L$, the eigenenergies are shifted downwards: in the limit
$L\to0_+$, the lowest eigenenergy tends to $-\infty$ as~$-1/L^2$, while
the higher eigenenergies tend to the $L=0$ values from above, with
corrections that are proportional to $L$~\cite{fewster-hydrogen}. Our
symmetric boundary condition on the lattice produces thus
eigenenergies that appear at $\lambda\to0$ to be in qualitative
agreement with the continuum eigenenergies at $L\to0$, although the
rate of convergence on the lattice is significantly slower.

The singularity resolution method appears therefore to be in
qualitative agreement with the continuum theory also for the symmetric
sector. The contrast between the symmetric sector and the
antisymmetric sector shows, however, that boundary conditions on the
quantum states at the singularity have a significant effect on the
spectrum even after the singularity has been regularized.

\section{Discussion}
\label{sec:discussion}

In this paper we have discussed polymer quantization in the $-1/|x|$
potential on the real line. We resolved the singularity of the
potential at $x=0$ by representing a derivative by its
finitely-differenced lattice version, by a technique that mimicks
the regularization of the inverse triad operator in
LQG~\cite{thomas}. Focusing on an equispaced lattice with a lattice
point at $x=0$, our numerical simulations indicated that the energy
eigenvalues in the antisymmetric sector converge rapidly to those of
the conventional continuum Coulomb problem as the lattice scale
$\lambda$ goes to zero. This is not unexpected, since antisymmetry on
the full real line corresponds to the conventional boundary condition
at the origin in the spherically symmetric three-dimensional Coulomb
problem. In contrast, for the symmetric sector we found that the
ground state eigenvalue appears to decrease without bound as $\lambda$
approaches zero, while the eigenvalues of the excited states appear to
approach the eigenvalues of the conventional Coulomb problem from
above. The singularity resolution method in the symmetric sector thus
yields dynamics that is qualitatively similar to that in the Coulomb
problem with an \emph{un\/}conventional choice of the self-adjointness
boundary condition at the origin, with $\lambda$ corresponding to the
length scale associated with this boundary condition.

We view these results as evidence that the singularity resolution
method yields physically reasonable results for the $-1/|x|$ potential
on the real axis, whether one regards the finite difference equation
simply as a discrete approximation to the Schr\"odinger dynamics in
$L_2(\mathbb{R})$ or whether one regards the lattice scale $\lambda$
as a fundamental length within polymer quantization. We emphasize that
the symmetric and antisymmetric sectors were found to have
qualitatively different spectra, where the antisymmetric sector
produces the conventional continuum eigenvalues in the limit of
small~$\lambda$. This shows that even after the singularity in the
potential has been regularized, boundary conditions that one may wish
to impose at the locus of the singularity can have a significant
effect on the spectrum.

We note in passing that our singularity-resolution technique may also
be of interest as a numerical technique in the context of pure
Schr\"odinger quantisation, as an alternative to numerical techniques
that invoke the asymptotic form of the continuum solution near the
singularity~\cite{gordon}. To explore this suggestion, one would need
to compare the convergence properties of our scheme, as evidenced in
Figures \ref{e0lam} and~\ref{e1lam}, to the convergence properties of
the matching scheme of~\cite{gordon}. 

Our results may be viewed as supporting Thiemann's regularization of
the inverse triad operator in LQG~\cite{thomas}. Furthermore, they
suggest that a boundary condition at the classical singularity may
have a significant role also in the loop quantum gravity context, both
when evolving through a spacelike singularity
\cite{mb1,ash-bojo-singularity,hajicek-unitary} and when setting
boundary conditions at a timelike singularity~\cite{horo-marolf}.

There are however at least three subtleties in this respect. (i) We
focused the numerical simulations on a regular lattice that has a
lattice point at the origin. Will the situation remain similar also on
irregular lattices, and is there a systematic control on the
singularity effects when the lattice is refined
\cite{fredenhagen-reszewski,corichi-vuka-zapata,helling-policastro}? 
In particular, would a significant symmetry-antisymmetry distinction
emerge also on irregular lattices? (ii) We introduced the polymer
quantization after first reducing the continuum Coulomb problem to the
spherically symmetric sector. If one wanted to discuss polymer
quantization corrections to the Coulomb energy levels from a
phenomenological viewpoint, the polymer Hilbert space should
presumably be introduced already at the level of three independent
spatial dimensions. (iii) The polymer Hilbert space utilized was the
genuine physical Hilbert space of the system, and the energy
eigenstates were simply the normalizable solutions to the eigenvalue
difference equation in the polymer inner product. In LQG, the polymer
Hilbert space is only the kinematical Hilbert space, and further
issues may emerge when the physical Hilbert space for solutions of the
Hamiltonian constraint is introduced
\cite{thiemann-book,lew-mar,GLMP}.

\bigskip

\noindent{\bf Acknowledgements}: 
We thank Edward Armour, Carsten Gundlach and Paul Matthews for helpful
discussions on the numerics. This work was supported in part by the
Natural Science and Engineering Research Council of Canada and by
PPARC (UK) Rolling Grant PP/D507358/1. JL~acknowledges hospitality and
financial support of the Isaac Newton Institute programme ``Global
Problems in Mathematical Relativity'' and of the Perimeter Institute
for Theoretical Physics.

\end{document}